\documentclass[twocolumn,aps,10pt,superscriptaddress]{revtex4-2}

\usepackage{amsmath}
\usepackage{amssymb}
\usepackage{amsfonts}
\usepackage{mathrsfs}
\usepackage{graphicx}

\newcommand{\DD}{\mathcal{D}}
\newcommand{\NN}{\mathcal{N}}
\newcommand{\rr}{\vec{r}}
\newcommand{\kk}{\vec{k}}

\newcommand{\kB}{k_{B}}
\newcommand{\TU}{T_{U}}
\newcommand{\NU}{N_{U}}
\newcommand{\GU}{u}
\newcommand{\NP}{N_{P}}
\newcommand{\hc}{\text{H.c.}}

\newcommand{\T}{\text{T}}

\newcommand{\dd}{\text{d}}

\newcommand{\LL}{{\mathcal L}}
\newcommand{\LLL}{\mathscr{L}}
\newcommand{\DDD}{\mathscr{D}}
\newcommand{\HH}{{\mathcal H}}
\newcommand{\bra}[1]{\langle#1|}

\newcommand{\ket}[1]{|#1\rangle}
\newcommand{\av}[1]{\bar{#1}}
\newcommand{\tr}{\text{tr}}
\newcommand{\ppp}{\\[3pt]}

\newcommand{\nppp}{\nonumber\\[3pt]}

\begin{document}

\title{Observable Unruh Effect and Unmasked Unruh Radiation}
\author{Charles H.-T. Wang}
\affiliation{Department of Physics, University of Aberdeen, Aberdeen AB24 3UE, United Kingdom}

\author{Gianluca Gregori}
\affiliation{Department of Physics, University of Oxford, Oxford OX1 3PU, United Kingdom}

\author{Robert Bingham}
\affiliation{Rutherford Appleton Laboratory, Didcot, OX11 0QX, United Kingdom}

\author{\\ Yakubu Adamu}
\affiliation{Department of Physics, University of Aberdeen, Aberdeen AB24 3UE, United Kingdom}

\author{Bethel N. Eneh}
\affiliation{Department of Physics, University of Aberdeen, Aberdeen AB24 3UE, United Kingdom}

\author{Ma\'e C. Rodriguez}
\affiliation{Department of Physics, University of Aberdeen, Aberdeen AB24 3UE, United Kingdom}

\author{Sarah-Jane Twigg}
\affiliation{Department of Physics, University of Aberdeen, Aberdeen AB24 3UE, United Kingdom}


\begin{abstract}
The Unruh effect, thereby an ideally accelerated quantum detector is predicted to absorb thermalized {\it virtual} photons and re-emit {\it real} photons, is significantly extended for laboratory accessible configurations. Using modern influence functional techniques, we obtain explicit expressions describing the excitation and relaxation of the quantum levels of an Unruh detector as a general noninertial open quantum system. Remarkably, for controllable periodical motions, an exact master equation is found for the Unruh detector within the prevailing framework of quantum optics with a well-defined Unruh temperature for given acceleration ($\alpha$), acceleration frequency ($\omega_\alpha$), and transition frequency ($\omega_0$) of the detector. We further show that the measurable Unruh temperatures and corresponding transition rates are comparable or higher than their values for the ideally accelerated cases if $c\omega_0$ and $c\omega_\alpha$ have similar orders of magnitude as $\alpha$. This allows us to select the transition rates of the detector to unmask Unruh radiation against Larmor radiation which has been a major competing noise. Our work suggests experiments with such settings may directly confirm the Unruh effect within the current technology, based on which a laboratory test of black hole thermodynamics will become possible.
\end{abstract}


\maketitle

\section{Introduction} The Unruh effect \cite{Unruh1976} is a surprising prediction from the quantum field theory \cite{Fulling1973, Davies1975} that an observer subject to a constant acceleration $\alpha$ relative to an inertial vacuum at zero temperature will experience a surrounding thermal bath at the Unruh temperature
\begin{eqnarray}
\TU(\alpha)
=
\frac{\hbar\alpha}{2\pi}
\label{TU}
\end{eqnarray}
in units with $c=1$ and $\kB=1$.

Furthermore, the accelerated observer is expected to evolve towards a thermal equilibrium with the environment at temperature $\TU(\alpha)$ by exchanging thermal photons, resulting in potentially measurable Unruh radiation. In this context, the accelerated observer is called an Unruh detector, which have excitable quantum levels \cite{DeWitt1979, Audretsch1994, Doukas2013}.
Alternatively, it has been suggested to observe the Unruh effect using accelerated electrons or positrons \cite{Chen1999, Brodin2008, Crowley2012, Lynch2021, Hegelich2022}.

Despite the QED origin of the Unruh effect, substantial interest in its direct detection has remained undamped \cite{Crispino2008}. This can be justified by the fundamental importance of Unruh radiation being parallel to the Hawking radiation \cite{Hawking1974} as a gateway to quantum gravity and further unified theories. The direct measurement of the Unruh radiation would also be a milestone for verifying the gravitational particle production process in the Early Universe \cite{Ford1987}.

While it is true that the difficulties of detecting Unruh radiation are largely due to experimental challenges, the limitations and indeed controversies of the existing theoretical descriptions of the Unruh effect have long added to the confusion of its measurement. Certain ambiguities in theoretical interpretations have even led some authors to believe that Unruh radiation does not exist \cite{Crispino2008}. There also appear to be perceptions that laboratory generated oscillatory accelerations lead to non-equilibrium conditions that could invalidate the notion of Unruh temperature \cite{Doukas2013}.

Here we provide a comprehensive and unambiguous theoretical framework for the Unruh effect based on the powerful influence functional technique in modern theory of open quantum systems with significant generalization to noninertial systems. This enable us to analyze an Unruh detector through the well-established quantum optics approaches with proven tools to reliably address the physical issues including the photon exchanges with the environment and the corresponding transition rates as well as conditions for relaxation to equilibrium. As a result we are able to derive the physically meaningful Unruh temperature for realistic laboratory conditions and unmask the resulting Unruh radiation with a careful treatment of relaxation times.

\section{Quantum detector as an open quantum system} Here we will derive a general master equation for a quantum detector coupled to the electromagnetic field in an environmental reservoir using the influence functional theory \cite{Breuer2002, Oniga2016}. This can be achieved by first considering the environment to consistent of a real massless scalar field $\phi$ described by the Lagrangian density
\begin{eqnarray}
\LL_R
=
-\frac12\,\eta^{\mu\nu}\phi_{,\mu}\phi_{,\nu}
\label{Lsc}
\end{eqnarray}
using the Minkowski metric $\eta_{\mu\nu}=$ diag$(-1,1,1,1)$
in a laboratory inertial frame.
The system of a quantum detector coupled to this environmental field will be statistically represented by a time-dependent total density operator (matrix) $\rho_T(t)$, satisfying the Liouville-von Neumann equation
\begin{eqnarray*}
\frac{\dd}{\dd t} \rho_T(t)
&=&
\int\dd^3 x\,\LLL \rho_T(t)
\ppp
&=&
-\frac i \hbar \int\dd^3 x\, [\HH(x), \rho_T(t)]
\end{eqnarray*}
where $\LLL$ is the total Liouville super operator, $\HH$ is the total Hamiltonian density, and $(x)=(t,\vec{x})$ denotes a spacetime point.
Provided the initial state $\rho_T(t_0)$, the above is formally solved by
\begin{eqnarray*}
\rho_T(t)
=
\T\exp\Big[\int_{t_0}^{t}\dd^4 x' \LLL(x') \Big] \rho_T(t_0)
\end{eqnarray*}
where $\T$ is the time ordering operator.

We consider a system-reservoir interaction Hamiltonian density of the form
\begin{eqnarray}
\HH_I
=
-\phi\, Q
\label{HHI}
\end{eqnarray}
with the corresponding Liouville super operator denoted by $\LLL_I$. Here the operator $Q(x)$ depends on the system variables with the assumed commutation relation $[Q(x),Q(x')]=0$ \cite{Breuer2002}.
Then the system density matrix $\rho=\tr_R(\rho_T)$ obtained by taking the partial trace over the environmental field which carries the reduced dynamic of the detector's degrees of freedom follows as
\begin{eqnarray}
\rho(t)
=
\T_S \Big\{
\tr_R \Big[ \T_\phi
\exp\Big(
\int_{t_0}^{t} \dd^4 x'\, \LL_I(x')
\Big) \rho(t_0) \Big]
\Big\}
\label{rhoeq}
\end{eqnarray}
in the interaction picture, where we have split the time ordering operator
$\T=\T_\phi \T_S$,
with $\T_\phi$ and $\T_S$ being the time ordering operators for the environmental and system variables respectively.
By virtue of the Wick's theorem \cite{Itzykson1980}, the identity
\begin{eqnarray}
&&\!\!\!\!\!\!\!\!\!\!\!\!\!
\T_S
\Big\{
[\LLL_I(x),\LLL_I(x')]\, \rho
\Big\}
\nppp
&=&
-\frac{1}{\hbar^2}
\T_S
\Big\{
[\phi(x), \phi(x')][Q(x) Q(x'), \rho]
\Big\}
\label{cmmm}
\end{eqnarray}
holds, as the operator $Q(x)$ at different times commute under $\T_S$ and the commutators of the environmental fields are $c$-numbers.

We consider a weak coupling so that the initial state of the system (at time $t_0$ can be given by a product state of the form $\rho_T (t_0) = \rho(t_0) \otimes \rho_R$, with $\rho(t_0)$ being the reduced density matrix of the matter at the initial time and $\rho_R$ being the reservoir density matrix, which is in a stationary state. The influence functional obtained from the time ordered exponential \eqref{rhoeq} can written as a cumulant expansion up to second order \cite{Breuer2002}, valid for any environment in a Gaussian state, including the vacuum and thermal states described by a quadratic Hamiltonian \cite{ferraro2005}.
Therefore, by substituting \eqref{cmmm} into \eqref{rhoeq}, we obtain the influence functional for the reduced open quantum system satisfying the non-Markovian master equation
\begin{eqnarray}
\frac{\dd\rho}{\dd t}
&=&
-\frac{1}{2\hbar}\int_{t_0}^{t}\! \dd t'
\dd^3 x\,\dd^3 x'
\Big\{
i \DD(x,x')
\big[Q(x), \{Q(x'), \rho\}\big]
\nppp&&
+
\NN(x,x')
\big[Q(x), [Q(x'),\rho]\big]
\Big\}
\label{disi}
\end{eqnarray}
using the dissipation and noise kernels
\begin{eqnarray*}
\DD(x,x')
&=&
\frac{1}{i\hbar}\,
\langle[\phi(x), \phi(x')]\rangle_R
\ppp
\NN(x,x')
&=&
\frac{1}{\hbar}\,
\langle  \{ \phi(x) , \phi(x') \} \rangle_R
\end{eqnarray*}
respectively, where $\langle\,\cdots\rangle_R$ denotes averaging over the reservoir, assumed to be stationary and in a Gaussian state.
This is the case if the environment is in thermal equilibrium at temperature $T$ with the Planck distribution $N(\omega)=\NP(\omega,T)$ given by
\begin{eqnarray}
\NP(\omega,T)
=
\frac1{e^{\hbar\omega/T}-1}.
\label{Pdist}
\end{eqnarray}

By incorporating vector and tensor components with suitable gauge symmetries, the master equation \eqref{disi} is readily generalized for the electromagnetic \cite{Breuer2002} and gravitational \cite{Oniga2016, Oniga2017} environments.
Specifically, by comparing the Hamiltonian densities of the electromagnetic and scalar fields, we see that the electromagnetic variant of \eqref{disi} is obtained with
\begin{eqnarray}
\phi
\to
A_{i}
,\,
Q
\to
j_{i}
,\,
\HH_I
\to
-A_{i}j_{i}
\label{Emap}
\end{eqnarray}
where $A_{i}$ is the vector potential $j_{i}$ is the electric current density \cite{Breuer2002}.
For completeness, by comparing the Hamiltonian densities of the linearized gravitational and scalar fields, we see that the gravitational variant of \eqref{disi} follows as
\begin{eqnarray}
\phi
\to
\frac{h_{ij}}{\sqrt{32\pi G}}
,\,
Q
\to
\sqrt{8\pi G}\,\tau_{ij}
,\,
\HH_I
\to
-\frac{1}{2}\,h_{ij}\tau_{ij}
\label{Gmap}
\end{eqnarray}
where $h_{ij}$ is the transverse-traceless (TT) gravitational wave amplitude and
$\tau_{ij}$ is the TT-stress tensor of a matter source \cite{Oniga2016, Oniga2017}.

\section{Moving quantum detector} An Unruh detector is by its nature on an accelerated trajectory. It is therefore necessary to extend the preceding formulation to allow for an arbitrary motion of a quantum detector that a laboratory test may execute. We have in mind an electron to be accelerated by an intense electric field to gain a large acceleration thanks to its high charge-to-mass ratio. This particle is also considered to be subject to a potential field that gives rise to quantum energy levels capable of being thermalized.

As a relatively simple model to capture these physical features, let us consider a point quantum particle having a uniform transition frequency $\omega_0$ between adjacent energy levels. This particle is assumed to move along a world line
$x(\tau)=(t(\tau),\rr(\tau))$
parameterized by the proper time $\tau$ \cite{Weinberg1972}
with a system-reservoir interaction Hamiltonian density \eqref{HHI} using
\begin{eqnarray}
Q(x)
=
\sqrt{\hbar}
\int\dd\tau\,
\delta^4(x-x(\tau))\,q(\tau)
\label{PARHI}
\end{eqnarray}
where $q(\tau)$ is a dimensionless Hermitian operator.
Here $Q(x)$ arises from taking current moments using \eqref{Emap} and physically can be related to the dipole of the particle-potential system as in a possible experiment scheme below.

Since we are considering a single transition frequency $\omega_0$, the time evolution of the $q(\tau)$ in the interaction picture is given by \cite{Breuer2002}
\begin{eqnarray}
q(\tau)
=
q^\dag\,e^{i\omega_0 \tau}
+
q\,e^{-i\omega_0 \tau}
\label{mPARq}
\end{eqnarray}
where
$q^\dag$ and $q$ act as raising and lowering operators for the particle's quantum energy states respectively.

We consider the zero temperature $T=0$ in the inertial laboratory frame, in which the late time quantum motion with $\tau \gg \tau_0$ in a coarse-grained time scale is described by the Markovian approximation of Eq. \eqref{disi} as follows
\begin{eqnarray}
\dot{\rho}
&=&
\frac{1}{16\pi^3}\!
\int_0^\infty\!\dd s\,
\Big\{
\big(
q\,\rho\,q^\dag
-
\rho\,q^\dag q
\big)
e^{-i\omega_0 s}
\nppp&&
+
\big(
q^\dag \rho\,q
-
\rho\,q\,q^\dag
\big)
e^{i\omega_0 s}
\Big\}
I(s,\tau)
+\hc
\label{accdisi4r0}
\end{eqnarray}
where
$\dot{\rho}={\dd\rho}/{\dd\tau}$,
$\hc$ denotes the Hermitian conjugate, and
%
\begin{eqnarray}
I(s,\tau)
&=&
\int\!\dd\Omega(\kk)\,
\int_0^\infty\!\dd\omega\,\omega
\nppp&&
\times\;
e^{i[\omega(t(\tau) - t(\tau-s))-\kk\cdot(\rr(\tau) - \rr(\tau-s))]}.
\label{Ifunc}
\end{eqnarray}
%


In particular, if $I(s,\tau)\approx I(s)$ over the coarse-grained time scale, then \eqref{accdisi4r0} takes the form
\begin{eqnarray}
\dot{\rho}
&=&
\frac{1}{16\pi^3}
\Big\{
\tilde{I}^+(\omega_0)
\big(
q\,\rho\,q^\dag
-
\rho\,q^\dag q
\big)
\nppp&&
+
\tilde{I}^+(-\omega_0)
\big(
q^\dag \rho\,q
-
\rho\,q\,q^\dag
\big)
\Big\}
+\hc
\label{accdisi4r}
\end{eqnarray}
in terms of the one-sided Fourier transform $\tilde{I}^+(\omega)$ of the coarse-grained $I(s)$, which is related to the full Fourier transform $\tilde{I}(\omega)$ by
\begin{eqnarray}
\tilde{I}^+(\omega)
=
\frac{1}{2}\,\tilde{I}(\omega)
+
\hbox{CP}
\label{I1w0Ar}
\end{eqnarray}
where CP denotes a certain Cauchy principal value.
It is well known that the Cauchy principal terms contribute only to the imaginary part of $\tilde{I}^+(\omega)$  related to a unitary evolution for $\rho$ which can be absorbed into a renormalized system Hamiltonian  \cite{Breuer2002}. Therefore, by introducing the dimensionless real ``Unruh function''
\begin{eqnarray}
U(\omega)
&=&
\frac{1}{16\pi^3|\omega|}\,
\tilde{I}(\omega)
\label{Ufunc}
\end{eqnarray}
Eq. \eqref{accdisi4r} becomes
\begin{eqnarray}
\dot{\rho}
=
\omega_0
\big\{
U(\omega_0)\,
\DDD[q]
+
U(-\omega_0)\,
\DDD[q^\dag]
\big\}\rho
\label{accdisi4r2}
\end{eqnarray}
in terms of the Lindblad super operator
%
\begin{eqnarray}
\DDD[q](\rho)
&=& 
q\,\rho\,q^\dag
-
\frac{1}{2}\,
\big\{
q^\dag q,\,\rho
\big\}.
\label{accdisi4US0}
\end{eqnarray}
Moreover, by introducing the ``Unruh distribution''
\begin{eqnarray}
\NU(\omega)
&=&
\frac{U(-\omega)}{U(\omega)-U(-\omega)}
\label{NU}
\end{eqnarray}
and the transition factor
\begin{eqnarray}
\GU(\omega)
&=&
U(\omega)-U(-\omega)
\label{GU}
\end{eqnarray}
we can cast \eqref{accdisi4US0} into a standard Lindblad form of the quantum optional master equation as follows
\begin{eqnarray}
\dot{\rho}
=
\omega_0\GU(\omega_0)
\big\{
(1+\NU(\omega_0))\, \DDD[q]
+
\NU(\omega_0)\, \DDD[q^\dag]
\big\}\rho.
\nppp
\label{accdisi4US}
\end{eqnarray}

The resulting quantum states evolve towards a completely decohered equilibrium state represented by a diagonal density matrix
\begin{eqnarray}
\rho
=
\sum_n \rho_{n}\ket{n}\bra{n}
\end{eqnarray}
with $0 \le \rho_{n} \le 1$ and $\sum_n \rho_{n}=1$. Furthermore, this equilibrium state is a thermal state with the temperature
\begin{eqnarray}
\TU(\omega)
=
\frac{\hbar\omega}{\ln[{U(\omega)}/{U(-\omega)}]}
\label{Tef}
\end{eqnarray}
which we shall refer to as the (generalized) Unruh temperature.

For example, if the quantum particle has harmonic quantum states, then we have
\begin{eqnarray}
\rho_n
=
(1-e^{-\hbar\omega_0/T})\,e^{-n\hbar\omega_0/T}
\label{accdisi4rhn2}
\end{eqnarray}
for $n=0,1,2,\ldots$, where $T=\TU(\omega_0)$.

\section{Quantum detector undergoing a constant acceleration} For a quantum detector under a constant acceleration $\alpha$, using \eqref{Ufunc} with $t(\tau)$ and $\rr(\tau)$ given by the Rindler coordinates \cite{Rindler1969}, we obtain the Unruh function to be
\begin{eqnarray}
U(\omega,\alpha)
=
\text{sgn}(\omega)(1+\NP(\omega,\TU(\alpha)))
\label{UNP}
\end{eqnarray}
%
where the Planck distribution given by \eqref{Pdist} evaluated for both signs of $\omega$ satisfying
$-\NP(-\omega,T)=1+\NP(\omega,T)$.
The corresponding transition factor $\eqref{GU}$ turns out to be unity.
Then using \eqref{UNP} we find that the Unruh distribution \eqref{NU} simply yields the Planck distribution
\begin{eqnarray}
\NU(\omega,\alpha)
&=&
\NP(\omega,\TU(\alpha))
\label{NU0}
\end{eqnarray}
given by \eqref{Pdist} in terms of the Unruh temperature \eqref{TU}. We have therefore recovered Unruh's result for a constant acceleration \cite{Unruh1976}.



\section{Quantum detector undergoing a periodic motion} Consider a periodic motion at oscillation frequency $\omega_\alpha$ in $\tau$ so that
$\rr(\tau)=\rr(\tau+\tau_\alpha)$ where $\tau_\alpha=2\pi/\omega_\alpha$ is the oscillation period. In terms of the time averaging of any quantity $p$ denoted by $\av{p}$ over an oscillation period $\tau_\alpha$, we can write
\begin{eqnarray*}
t(\tau)
=
\av{\gamma} \tau + \xi(\tau)
\end{eqnarray*}
where
$\gamma = \dd t/\dd \tau$
is the Lorentz factor for the particle motion and $\xi(\tau)=\xi(\tau+\tau_\alpha)$ is some function as the periodic variation of the laboratory inertial time relative to the proper time of the particle. For simplicity, let us work in units with $\omega_\alpha = 1$ and hence $\tau_\alpha = 2\pi$.
Then \eqref{Ifunc} becomes
\begin{eqnarray*}
I(s,\tau)
=
\int_0^\infty\dd\omega\,
\omega\,f(s,\omega,\tau)\,
e^{i\,\av{\gamma}\,\omega s}
\end{eqnarray*}
where
\begin{eqnarray*}
f(s,\omega,\tau)
=
\!\int\!\dd\Omega(\kk)\,
e^{i[
\omega(\xi(\tau)-\xi(\tau-s))
-
\kk\cdot(\rr(\tau)-\rr(\tau-s))
]}.
\end{eqnarray*}
It follows from the periodicity of $\xi(\tau)$ and $\rr(\tau)$ that the real and imaginary parts of $\av{f}(s,\omega)$ are even and odd functions of $s$ respectively.
This results in the Fourier coefficients $f_n(\omega)$ of $\av{f}(s,\omega)$ to be real for all integers $n$. Consequently, the Unruh function introduced in \eqref{Ufunc} so that the quantum optical equation \eqref{accdisi4r2} holds, now takes the form
\begin{eqnarray}
U(\omega)
=
\frac{1}{4\pi|\omega|\av{\gamma}^2}\,
\sum_{n < \omega}
(\omega - n)\,
f_n\Big(\frac{\omega - n}{\av{\gamma}}\Big)
\label{I1w0Ar2b}
\end{eqnarray}
where $f_n(\omega)$ are the Fourier coefficients of $\bar{f}(s,\omega)$
over an interval of $s$ of length $2\pi$.

\section{Quantum detector in a circular motion} As a special case of periodic motions, we can consider a circular motion with
\begin{eqnarray*}
\rr(\tau)
&=&
\frac{\alpha}{\omega_\alpha^2}\,
(\cos(\omega_\alpha \tau),\sin(\omega_\alpha \tau),0)
\end{eqnarray*}
having a centripetal acceleration $\alpha$, rotational frequency $\omega_\alpha$, and
the constant Lorentz factor
$\gamma=\sqrt{1+{\alpha^2}/{\omega_\alpha^2}}$.
In units with $\omega_\alpha=1$, Eq. \eqref{I1w0Ar} then becomes
\begin{eqnarray*}
I(\omega_0)
=
2\int_0^\infty\!\dd\omega
\int_0^\infty\!\dd s\, \omega  e^{-i(\omega_0 - \gamma \omega) s}
J_0\left(2\alpha\omega\sin\frac{s}{2}\right).
\end{eqnarray*}
This leads to the Unruh function
\begin{eqnarray*}
U(\omega,\alpha)
=
\frac{1}{2\pi|\omega|(1+\alpha^2)}\!
\sum_{n < \omega}
(\omega - n)\, d_n\Big(\frac{\omega - n}{\sqrt{1+\alpha^2}}\Big)
\end{eqnarray*}
in terms of
\begin{eqnarray}
d_n(\omega)
&=&
\int_{0}^{2\pi}\!
J_0\left(2 \alpha\omega \sin\frac{s}{2}\right)\cos ns
\,\dd s
\label{I1w0Arotjn}
\end{eqnarray}
from which the corresponding Unruh temperature \eqref{Tef}, distribution \eqref{NU}, and transition factor \eqref{GU} are readily evaluated.

\section{Possible detection of unmasked Unruh radiation} Here we consider the Unruh radiation of an accelerated electron subject to a potential. As an Unruh detector, the quantum states associated with the potential should be independent of the motion and this will be the case when the potential is induced by a uniform magnetic field in the direction of a linear acceleration.

We shall demonstrate that the Unruh radiation of such a system is a direct consequence of the coupling between the Landau levels of the electron and environmental photons even at zero temperature.

\begin{figure}[!t]
\includegraphics[width=0.8\linewidth]{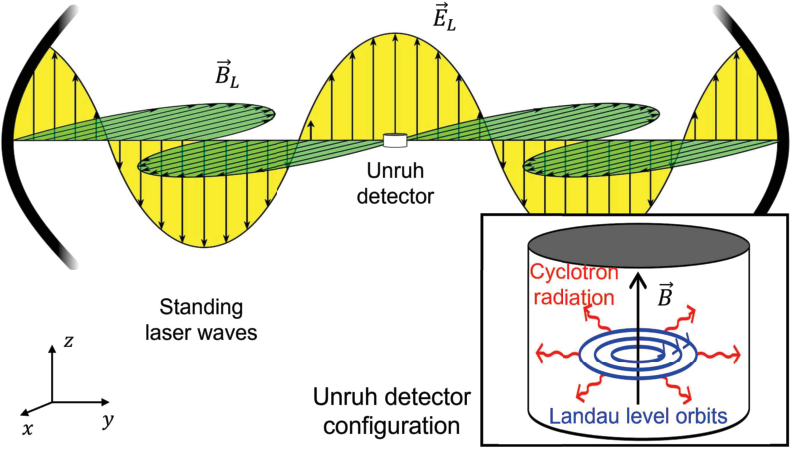}
\caption{Schematic diagram of Unruh radiation as thermal cyclotron radiation through quantum Landau-level transitions of a magnetically confined electron accelerated by laser.}
\label{fig-1}
\end{figure}

As illustrated in Fig. \ref{fig-1}, we start by considering the spontaneous emission from an electron (with charge $e$ and mass $m$) in a uniform magnetic field $\vec{B}$ along the $z$-direction with a cyclotron frequency
$\omega_c = B e/m$ . We will then consider the effect of an oscillatory linear acceleration in the $z$-direction.

The Landau quantization of the electron takes place on the transverse plane with coordinates $x$ and $y$ and are shifted by the ladder operators given in terms of the dimensionless variables
$X=x/x_c$ and $Y=y/x_c$ with $x_c=\sqrt{\hbar/m \omega_c}$, by the expressions
\begin{eqnarray*}
a
&=&
\tfrac{1}{\sqrt{2}}\,\big[
\big(
\tfrac{X}{2}
+
\partial_X
\big)
-
i\big(
\tfrac{Y}{2}
+
\partial_Y
\big)
\big]
\ppp
b
&=&
\tfrac{1}{\sqrt{2}}\,\big[
\big(
\tfrac{X}{2}
+
\partial_X
\big)
+
i\big(
\tfrac{Y}{2}
+
\partial_Y
\big)
\big]
\end{eqnarray*}
satisfying the nontrivial commutation relations
\begin{eqnarray*}
[a , a^\dag]
=
[b , b^\dag]
= 1.
\end{eqnarray*}

However, the operators $b$ and $b^\dag$ only shift the Landau level orbits on the transverse plane as the system Hamiltonian of the electron takes the form
\begin{eqnarray*}
H_S
&=&
\hbar\omega_c \big(a^\dag a + \tfrac{1}{2}\big).
\end{eqnarray*}

The coupling of the electron Landau levels to the radiation field under the dipole approximation is through interaction Hamiltonian
\begin{eqnarray*}
H_I
&=&
-\vec{D}\cdot\vec{E}_R
\end{eqnarray*}
where $\vec{E}_R$ is the electric field operator of the radiation field and
$\vec{D}= e (x, y, 0)$ is the dipole operator of the orbiting electron.
Following the theoretical framework laid out in preceding sessions, we find that the cyclotron configuration as an open quantum system at temperature $T$ to satisfy the master equation
\begin{eqnarray}
\dot{\rho}
=
\Gamma_c
\big\{
(1 + N(\omega_c))\,\DDD[a]
+
N(\omega_c)\,\DDD[a^\dag]
\big\}\rho
\label{maseqa}
\end{eqnarray}
in terms of the Planck distribution $N(\omega_c)=\NP(\omega_c,T)$ and the spontaneous decay rate
\begin{eqnarray}
\Gamma_c
=
\frac{e^2\omega_c^2}{3\pi\epsilon_0 m c^3}
\label{Gamc}
\end{eqnarray}
in full units.
In the classical limit, Eqs. \eqref{maseqa} and \eqref{Gamc} correctly recovers the Larmor formula for the acceleration radiation from a circularly orbiting charge.

When the electron is additionally accelerated by the electric field $\vec{E}_L$ of a standing high-power laser wave in the direction of $\vec{B}$ at the zero temperature, the vacuum fluctuations of the electromagnetic field are thermalized in the cyclotron's moving frame, effecting the above spontaneous emission into thermal radiation manifested as Unruh radiation, described by the master equation \eqref{maseqa} with the following modifications
\begin{eqnarray*}
\Gamma_c
&\to&
\Gamma_U =
\Gamma_c\,\GU(\omega,\alpha),
\ppp
N(\omega)
&\to&
\NU(\omega,\alpha),
\ppp
T
&\to&
\TU(\omega,\alpha).
\end{eqnarray*}

\begin{figure}[!t]
\includegraphics[width=0.65\linewidth]{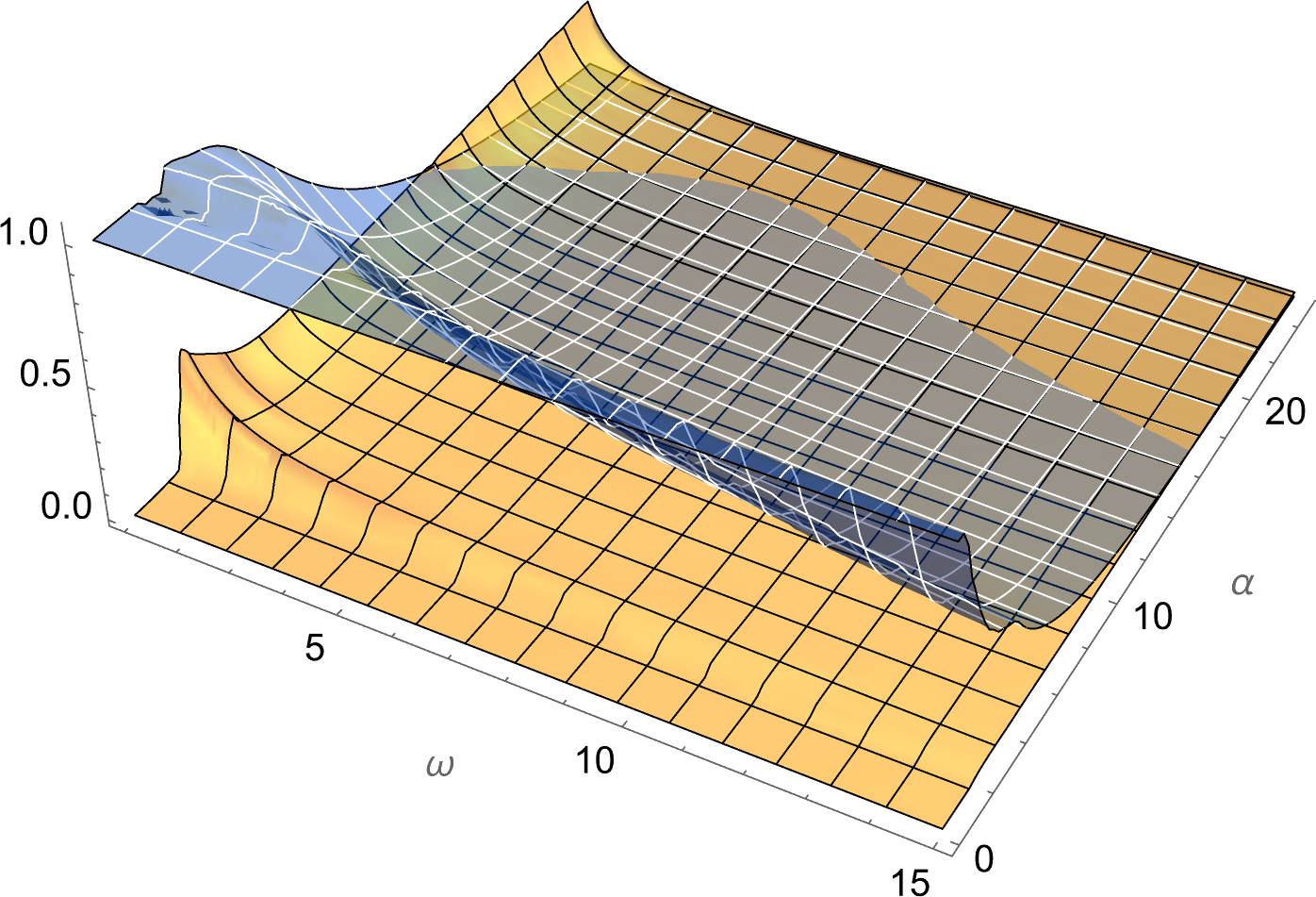}
\caption{The Unruh distribution $\NU(\omega_c,\alpha_L)$ (amber) and transition factor $\GU(\omega_c,\alpha_L)$ (blue) for the oscillating motion. Here $\omega=\omega_c/\omega_L$ and $\alpha=\alpha_L/c\omega_L$. This factor is around unity if both $\alpha$  and $\omega$ are order one but falls off to zero rapidly.}
\label{fig-2}
\vspace{11pt}
\includegraphics[width=0.65\linewidth]{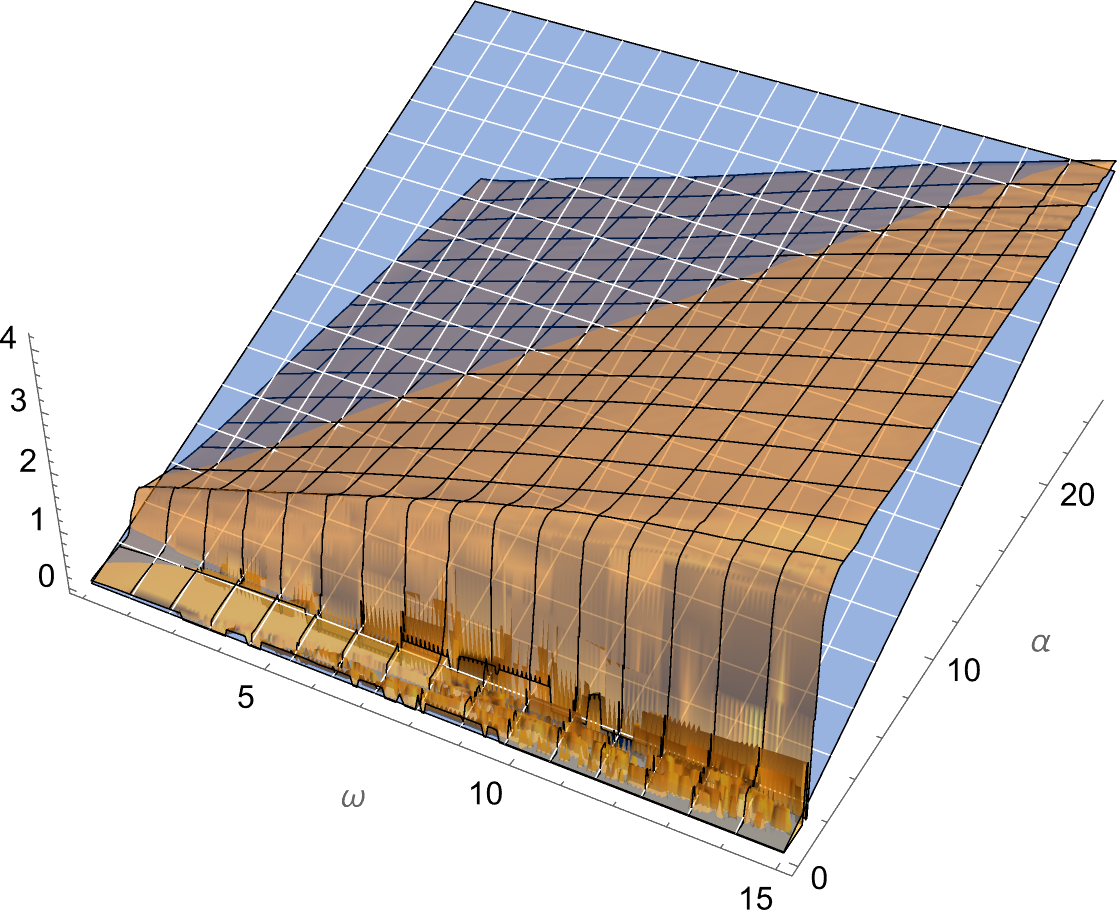}
\caption{The Unruh temperature $\TU(\omega_c,\alpha_L)$ (amber) with an oscillatory acceleration amplitude $\alpha_L$ compared with $\TU(\alpha_L)$ (blue) with a uniform acceleration $\alpha_L$. Here $\omega=\omega_c/\omega_L$, $\alpha=\alpha_L/c\omega_L$, and both $\TU(\omega_c,\alpha_L)$ and $\TU(\alpha_L)$ are evaluated in units of $\hbar\omega_L$.}
\label{fig-3}
\end{figure}

As shown in Figs. \ref{fig-2} and \ref{fig-3}, the Unruh temperatures and corresponding transition rates are comparable or higher than their values for the uniformly accelerated cases if $c\omega_0$ and $c\omega_L$ have similar orders of magnitude as $\alpha_L$.

For example, with the achievable magnetic field $B = 1000$ T \cite{Nakamura2018}, the accelerated cyclotron system has a transition frequency $\omega_0$ given by the cyclotron frequency
$\omega_c = 1.76\times10^{14}$ rad/s
with the transition/decay rate $\Gamma_c = 3.88\times 10^{5}$ s$^{-1}$.
Using the available high-power laser acceleration $\alpha_L = 6.0 \times 10^{23}$ m/s$^2$ at the frequency $\omega_L = 2\pi \times 3.0 \times 10^{13}$ rad/s \cite{Haberberger2010}, we can achieve the Unruh temperature
$\TU(\omega_c,\alpha_L) = 852$ K with the transition rate
$\Gamma_U = 1.95\times 10^{5}$ s$^{-1}$ corresponding to the Unruh detector relaxation time $\tau_U = 5.12 \mu$s. Since this relaxation time $\tau_U$ is much larger than the QED time scale, it is possible to detect the persistent Unruh radiation within the interval $\tau_U$ from the moment when the laser is switched off which discontinues masking radiations including the Larmor and other forms of acceleration radiations \cite{Soda2022}.

\section{Conclusion} The Unruh effect is an exceptional prediction in physics in that it is both widely celebrated and debated. We have provided a ground-up derivation of the Unruh effect with minimum assumptions, going far beyond its usual remit related to an idealized constant acceleration of a detector and have extended the scope of the effect to real-world scenarios with variable accelerations for finite durations while keeping the crux of the effect -- the physical reality of the Unruh temperature and radiation. To guide experimental tests, we have demonstrated that the measurement of the Unruh effect are feasible inside certain windows of the laboratory parameter space, which are previously unnoticed but are well within the current technology.

\vspace{-11pt}\acknowledgments

The authors are grateful to the Cruickshank Trust, Scotland (C. W.), the TETFund, Nigeria (Y. A. and B. E.), and the Carnegie Trust, Scotland (M. R.) for financial support.



\begin{thebibliography}{99}

\bibitem{Unruh1976}
W. G. Unruh,
Notes on black-hole evaporation,
Phys. Rev. D 14, 870 (1976).

\bibitem{Fulling1973}
S. A. Fulling,
Nonuniqueness of Canonical Field Quantization in Riemannian Space-Time,
Phys. Rev. D 7, 2850 (1973).

\bibitem{Davies1975}
P. C. W. Davies,
Scalar production in Schwarzschild and Rindler metrics,
J. Phys. A 8, 609 (1975).

\bibitem{DeWitt1979}
B. S. DeWitt,
Quantum gravity: the new synthesis,
in {\it General Relativity: An Einstein Centenary Survey}, pp. 680,
ed. S. W. Hawking, and W. Israel,
(Cambridge U. P., Cambridge, 1979).

\bibitem{Audretsch1994}
J. Audretsch and R. M\"uller,
Spontaneous excitation of an accelerated atom: The contributions of vacuum fluctuations and radiation reaction,
Phys. Rev. A 50, 1755 (1994).

\bibitem{Doukas2013}
J. Doukas, S.-Y. Lin, B. L. Hu, R. B. Mann,
Unruh Effect under non-equilibrium conditions: Oscillatory motion of an Unruh-DeWitt detector,
JHEP 11, 119 (2013.)

\bibitem{Chen1999}
P. Chen and T. Tajima,
Testing Unruh radiation with ultraintense lasers,
Phys. Rev. Lett. 83, 256 (1999).

\bibitem{Brodin2008}
G. Brodin, M. Marklund, R. Bingham, J. Collier, and R. G. Evans,
Laboratory soft x-ray emission due to the Hawking-Unruh effect?
Class. Quantum Grav. 25, 145005 (2008).

\bibitem{Lynch2021}
M. H. Lynch et al.,
Experimental observation of acceleration-induced thermality,
Phys. Rev. D 104, 025015 (2021).


\bibitem{Crowley2012}
B. J. B. Crowley, R. Bingham, R. G. Evans et al.,
Testing quantum mechanics in non-Minkowski space-time with high power lasers and 4th generation light sources,
Sci. Rep. 2, 491 (2012).

\bibitem{Hegelich2022}
B. M. Hegelich, L. Labun, O. Z Labun et al.,
Electron response to radiation under linear acceleration;  Classical, QED, and accelerated frame predictions,
Phys. Rev. D 105, 096034 (2022).

\bibitem{Crispino2008}
L. C. B. Crispino, A. Higuchi, and G. E. A. Matsas,
The Unruh effect and its applications,
Rev. Mod. Phys. 80, 787 (2008)
and references therein.

\bibitem{Hawking1974}
S. W. Hawking,
Black hole explosions?
Nature 248, 30 (1974).

\bibitem{Ford1987}
L. H. Ford,
Gravitational particle creation and inflation,
Phys. Rev. D 35, 2955 (1987)

\bibitem{Itzykson1980}
C. ltzykson and J.-B. Zuber,
{\it Quantum Field Theory},
(McGraw-Hill, New York, 1980).

\bibitem{Rindler1969}
W. Rindler,
{\it Essential Relativity}
(Van Nostrand Reinhold Co, New York, 1969).

\bibitem{Weinberg1972}
S. Weinberg,
{\it Gravitation and Cosmology},
(Wiley, New York, 1972).

\bibitem{Breuer2002}
H.-P. Breuer and F. Petruccione,
{\it The Theory of Open Quantum Systems}
(Oxford University Press, New York, 2002)
and references therein.

\bibitem{ferraro2005}
A. Ferraro, S. Olivares and M. G. A. Paris, {\it Gaussian States in Continuous Variable Quantum Information} (Bibliopolis, Napoli, 2005).

\bibitem{Oniga2016}
T. Oniga, C. H.-T. Wang, 
Quantum gravitational decoherence of light and matter, 
Phys. Rev. D 93, 044027 (2016).

\bibitem{Oniga2017}
T. Oniga, C. H.-T. Wang, 
Quantum coherence, radiance, and resistance of gravitational systems,
Phys. Rev. D 96, 084014 (2017).

\bibitem{Nakamura2018}
D. Nakamura, A. Ikeda, H. Sawabe, et al.,
Record indoor magnetic field of 1200 T generated by electromagnetic flux compression,
Rev. Sci. Instrum. 89, 095106 (2018).

\bibitem{Haberberger2010}
D. Haberberger, S. Tochitsky, and C. Joshi,
Fifteen terawatt picosecond CO$_2$ laser system,
Optics Express 18, 17865 (2010).

\bibitem{Soda2022}
B. \v{S}oda, V. Sudhir, and A. Kempf
Acceleration-Induced Effects in Stimulated Light-Matter Interactions,
Phys. Rev. Lett 128, 163603 (2022).


\end{thebibliography}
\end{document}